\documentclass{article}
\usepackage{fancyheadings}
\usepackage[british]{babel}
\usepackage{cite}

\usepackage{amsmath}
\usepackage{amssymb}

\begin{document} 

\pagenumbering{arabic}
\title{A note on dilaton gravity with non-smooth potentials}
\author{Christian G. B\"ohmer\footnote{e-mail: {\tt boehmer@hep.itp.tuwien.ac.at}}\\
        \footnotemark[1]~The Erwin Schr\"odinger International
        Institute for Mathematical\\ Physics,
        Boltzmanngasse 9, A-1090 Wien, Austria.
        \and
        \footnotemark[1]~Institut f\"ur Theoretische Physik, 
        Technische Universit\"at Wien,\\ 
        Wiedner Hauptstrasse 8-10, A-1040 Wien, Austria.
        \and
        Piotr Bronowski\footnote{e-mail: {\tt pbronowski@poczta.fm}}\\
        \footnotemark[2]~Institute for Theoretical Physics, 
        University of Lodz,\\ 
        Pomorska 149/153, PL-90-236 Lodz, Poland.}
\date{}
\maketitle

\thispagestyle{fancy}
\setlength{\headrulewidth}{0pt}
\rhead{Preprint ESI 1602, TUW--05--03}

\begin{abstract}
Recent interest in brane world models motivates the investigation
of generic first order dilaton gravity actions, with potentials
having some non-smoothness. We consider two different types of
$\delta$-like contributions in the action and analyse their
effects on the solutions. Furthermore a second
source of non-smoothness arises due to the remaining ambiguities 
in the solutions in the separated smooth patches,
after fixing all other constants by matching and asymptotic conditions.
If moreover staticity is assumed, we explicitly construct exact solutions.

With the methods described, for example models with
point like sources or brane world models (where the second 
source of non-smoothness becomes crucial),
can now be treated as non-smooth dilaton gravity theories.
\end{abstract}
\begin{center}
Dedicated to Wolfgang Kummer at the occasion of his {\it Emeritierung}
\end{center}
\mbox{} 
\mbox{} \\
\mbox{} 

\newpage
\section{Introduction}

In recent years brane world scenarios~\cite{Randall:1999vf} like e.g.~brane
world cosmology~\cite{Brax:2004xh} have been widely investigated
as alternatives to the standard model of particle physics. In terms
of the metric such models are realised by a continuous metric which
has a discontinuity in the first derivative. In the action this is represented
by $\delta$-like `matter' terms. Usually the resulting gravitational field
equations were analysed in the standard second order formulation,
see e.g.~\cite{Mann:1990gh,Melis:2004bk} for a 
$\delta$-like `matter' term in the 
second order dilaton gravity action. Here we present the necessary 
considerations to analyse these models in the first order formalism 
of dilaton gravity.

Lower and two-dimensional gravity models, cf~\cite{Brown:1988,Nojiri:2000ja,Grumiller:2002nm}, 
have become a rich source of investigation in the last 
two decades since they allow solutions to
problems, not yet solved in the four-dimensional case, like the
background independent quantisation of gravity in $1+1$ dimensions.
The two-dimensional Einstein-Hilbert action is only a surface term,
the Gauss-Bonnet term, hence intrinsically two-dimensional models
must be supplemented by further structure since they are otherwise
locally trivial. In compactifications from higher dimensions this
additional structure is provided automatically by the dilaton field. For spherically
reduced four-dimensional gravity the dilaton field can be regarded 
as the surface area of the sphere. 

The resulting second order action can be written in an equivalent first 
order form by introducing auxiliary fields~\cite{Schaller:1994es}.
Its action in two dimensions~\cite{Grumiller:2002nm,Schaller:1994es} reads
\begin{multline}
      L=\int \Bigl[
      X^+(d-\omega)\wedge e^- + X^-(d+\omega)\wedge e^+\\ + Xd\omega +
      V(X)\epsilon + X^+X^- U(X)\epsilon +{\mathcal L}_m \epsilon \Bigr] \,,
      \label{eq:note1}
\end{multline}
where $X$ is the dilaton field, $e^a$ is the zweibein 1-form,
$\epsilon=e^+\wedge e^-$ is the volume 2-form. The 1-form $\omega$ represents the 
spin-connection $\omega^a{}_b=\epsilon^a{}_b\omega$ with $\epsilon_{ab}$ 
being the totally skew-symmetric Levi-Civit{\'a} symbol, $\mathcal{L}_m$
stands for some matter Lagrangian. Note that the action~(\ref{eq:note1}) 
depends on two auxiliary fields $X^{\pm}$. 

The dilaton potential $\mathcal{V}(X,X^+X^-)$ 
is an arbitrary function depending solely on Lorentz 
invariant combinations of $X$ and $X^+X^-$. 
All physically relevant models are covered by a special form
$\mathcal{V}(X,X^+X^-)=V(X)+X^+X^-U(X)$.
In the case of spherical symmetry (or more precisely, spherical
reduction of spherically symmetric four-dimensional gravity)
$V(X)=-\lambda^2$ and $U(X)=-1/(2X)$. The parameter $\lambda$ of 
mass dimension one arises as a scaling factor of the dilaton.

If $\mathcal{V}(X,X^+X^-)$ contains non-smooth contributions, 
the classical solution for such a scenario 
can be obtained straightforwardly by solving the model in patches where the 
potential is smooth and by imposing as matching condition continuity of the 
Killing norm or equivalently continuity of the metric. This, together with 
asymptotic conditions on the line element, 
will fix all integration constants following standard methods 
(chapter~3 of~\cite{Grumiller:2002nm}).

\section{Non-smooth potentials}

The equations of motion derived from the action~(\ref{eq:note1})
by varying with respect to $\omega$, $e^{\mp}$, $X$ and $X^{\mp}$
are given by
\begin{align}
      &dX+X^-e^+ - X^+e^-=0,
      \label{eq:neom1} \\
      &(d\pm\omega)X^{\pm}\mp{\cal V}e^{\pm}=0,
      \label{eq:neom2} \\
      &d\omega+\epsilon\frac{\partial \mathcal{V}}{\partial X}=0,
      \label{eq:neom3} \\
      &(d\pm\omega)\wedge e^\pm+\epsilon\frac{\partial \mathcal{V}}{\partial X^\mp}=0,
      \label{eq:neom4}
\end{align}
respectively. The matter parts $W^{\pm}=\delta L_m/\delta e^{\mp}$ 
and $W=\delta L_m/\delta X$ were neglected because in this paper
we will only consider matter contributions that can be incorporated
into the dilaton potentials. Therefore, one may argue that matter
contributions are not necessary since the sources of non-smoothness
are only the dilaton potentials. However, the interpretation in
terms of matter is often useful.

Linear combination of the two equations~(\ref{eq:neom2})
with coefficients $X^{\mp}$, respectively, together
with~(\ref{eq:neom1}) yields
\begin{align}
      d(X^+X^-)+X^+X^- U(X)dX+V(X)dX=0.
\end{align}
Multiplying this by the integrating function
\begin{align}
      I(X)=c \exp\int^X U(y)dy,
      \label{eq:I}
\end{align}
one obtains the constant of motion (Casimir function)
\begin{align}
      d\mathcal{C}=0,\qquad \mathcal{C}=IX^+X^-+w,
\end{align}
where $w$ is defined by
\begin{align}
       w(X)=\int^X I(y)V(y)dy.
      \label{eq:note6}
\end{align}
The multiplicative ambiguity in the definition of the 
integrating factor $I(X)$ ($c\in\mathbb{R}$) has been
exhibited explicitly for later convenience.

If some discontinuities in the potentials $U(X)$ and $V(X)$ is allowed, 
then the smoothness of all other geometrical quantities can be 
read off from the equations of motion~(\ref{eq:neom1})--(\ref{eq:neom4}), 
see table~\ref{tab}. We introduce the notion of a 
smoothness degree $n$ which refers to $C^n$, e.g.~a 
continuous but non-differentiable function has smoothness degree of 0.
Only non-smoothness degrees $n,m\geq-2$ are considered, i.e.~no 
$\delta'$-like terms in the action will be allowed.
The superscript in $\mathcal{C}^{(g)}$ indicates the conserved quantity
with respect to geometry. If additional matter fields are present, the
total conserved quantity is $\mathcal{C}^{(tot)}=\mathcal{C}^{(g)}+\mathcal{C}^{(m)}$. 
The case where $m=n=-2$ in general is ill defined, since 
the integrand of $w$, namely $I(X)V(X)$, may not be a well defined 
distribution. If we require a continuous line element, at first glance, 
there can be only discontinuities in the dilaton potential, but no 
distributional contributions. However, we are still free to choose 
Casimir constants in each patch separately, and in this way remove such 
discontinuities of the metric. On the other hand, the smoothness degree
in some special cases can be improved, note e.g $|x|\frac{d|x|}{dx}=x$.
\begin{table}[ht]
\begin{center}
\begin{tabular}{|l||c|c|} 
\hline
Quantity & {\rm Symbol} & {\rm naive smoothness-degree} \\ \hline \hline
potential & $U$ & $m\in\mathbb{Z}$ \\ 
potential & $V$ & $n\in\mathbb{Z}$ \\ 
integrating factor & $I$ & $m+1$ \\
function & $w$ & min($m+2,n+1$) \\
conserved quantity & $\mathcal{C}^{(g)}$ & min($m+2,n+1$) \\
dual basis & $e^\pm$ & min($m+1,n+1$) \\
spin connection & $\omega$ & min($m,n$) \\
metric & $g_{\mu\nu}$ & min($m+1,n+1$) \\
curvature & $R$ & min($m-1,n-1$) \\
torsion & $T$ & min($m,n$) \\
\hline
\end{tabular}
\end{center}
\caption{Smoothness degree of the dilaton potentials and quantities derived thereof.}  
\label{tab}
\end{table}

For generic potentials $V(X)$ and $U(X)$ the full
solution to the equations of motions derived from
the action~(\ref{eq:note1}) reads~\cite{Klosch:1996fi,Grumiller:2002nm}
\begin{align}
      &e^+=X^+Idf,\qquad e^-=\frac{dX}{X^+}+X^-Idf,
      \label{eom1}\\
      &\omega=-\frac{dX^+}{X^+}+\mathcal{V}I df,
      \label{eom2}\\
      &\mathcal{C}=IX^+X^-+w={\rm const.},
      \label{eom3}\\
      &ds^2 = 2Idf\otimes dX + 2I(\mathcal{C}-w)df\otimes df.
      \label{eom4}
\end{align}
In the following some $\delta$-like contributions $\delta(X-X_0)$
to the potentials $V(X)$ and $U(X)$ will be introduced:

(i)~Let us start with an additional contribution to the
dilaton gravity action~(\ref{eq:note1}) of the form $F(X)\delta(X-X_0)$,
which can be absorbed in the dilaton potential $V(X)$,
by noting $\tilde{V}(X)=V(X)+F(X)\delta(X-X_0)$. Without loss 
of generality we may assume that $X_0$ is positive. Together 
with a generic potential $U(X)$ the complete solution 
to the equations of motions is given by (\ref{eom1})--(\ref{eom4}). 
The effect of such a $\delta$-like term to the potential $V(X)$
enters the solution in the absolutely conserved quantity (\ref{eom3}) 
as
\begin{align}
      \mathcal{C}=I(X)X^+X^- + \tilde{w}(X),
\end{align}
where
\begin{align}
      \tilde{w}(X)&=\int^X I(y)V(y)dy +\int^X F(y)\delta(y-X_0)dy\\
                  &=\begin{cases}
		          w(X) + F(X_0), & X>X_0,\\
		          w(X), & X<X_0.
		    \end{cases}
\end{align}
For a continuous line element~(\ref{eom4})
at $X=X_0$ the two Casimir constants $\mathcal{C}_{X>X_0}$ and $\mathcal{C}_{X<X_0}$
of the respective patches $X>X_0$ and $X<X_0$ are related by
 $\mathcal{C}_{X>X_0}=\mathcal{C}_{X>X_0}+F(X_0)$.
So the matching condition, applied to the Killing norm,  
gives a shift between the two Casimir constants in
the two different patches. Note that in this case the metric also is
differentiable at $X=X_0$ (in fact $C^{\infty}$). If the metric's first derivative
shall jump, this can only be achieved by introducing some
non-smoothness via the relation between the dilaton and the 'physical' 
coordinate of the manifold, see the discussion below.

(ii)~Secondly, we consider an additional term to the 
action of the form $X^+X^-G(X)\delta(X-X_0)$,
which now can be absorbed in the other dilaton potential $U(X)$,
$\tilde{U}(X)=U(X)+G(X)\delta(X-X_0)$. This yields
\begin{align}
      \tilde{I}(X)&=c\exp\Bigl(\int^X U(y)+\int^X G(y)\delta(y-X_0)dy\Bigr)\\
                  &=\begin{cases}
		          I(X)G_0, & X>X_0,\\
		          I(X), & X<X_0,
		    \end{cases}
\end{align}
where we denoted $G_0=\exp(G(X_0))$ and where $\tilde{w}(X)$ 
is given by
\begin{align}
      \tilde{w}(X) = \int^X \tilde{I}(y)V(y)dy
                   = \begin{cases}
		          w(X)G_0, & X>X_0,\\
		          w(X), & X<X_0.
		      \end{cases}
\end{align}
The motivation of the term $X^+X^-G(X)\delta(X-X_0)$ comes
from the following observation. $\delta$-like contributions 
in the equivalent second order formulation of dilaton 
theories~\cite{Grumiller:2002nm} are usually of the form $\delta(r-r_0)$
which with $X=X(r)$ can be written as $\delta(r-r_0)=|X'(r)|\delta(X(r)-X_0)$.
Equation (\ref{eq:neom1}) expresses the directional derivatives of $X$ 
in terms of $X^\pm$ and therefore the term $X'$ naturally suggests 
then the above mentioned form. For a discussion of distributions
on general manifolds, see e.g.~\cite{Balasin:1993}. The term $|X'(r_0)|$
can also be regarded as the Jacobian of the coordinate transformation
between the dilaton $X$ and the `physical' coordinate $r$.

As already discussed in the introduction the metric should be continuous
at $X=X_0$. For the second case this yields 
$\mathcal{C}_{X<X_0}=\mathcal{C}_{X>X_0}G_0-w(X_0)(1-G_0^2)$.
In contrast to the previous case the metric is not differentiable
at the point $X=X_0$.

To specify the gauge of the metric to Eddington-Finkelstein form, we
introduce the new variable $r$ by $dX=I^{-1}dr$,
\begin{align}
      ds^2 &= 2df\otimes dr + 2I(\mathcal{C}-w)df\otimes df. 
      \label{eq:ef}
\end{align}
This coordinate redefinition may be an additional source of non-smoothness of
the metric. With the help of conformal transformations,
one can always eliminate the potential $U$, i.e. get a conformally 
related dilaton gravity model with  $U=0$, (cf~\cite{Grumiller:2002nm}),
which of course corresponds to a different theory.

In this case the integrating factor $I$ is a constant, therefore
one should have a closer look at these situations.
The coordinate $r$ and the dilaton $X$ are then
related by $X=I^{-1}r+X_0$, where $X_0$ is a constant of integration.
This relation allows the situation $X=0$ for $r=-IX_0$. The simplest way
to avoid such singular points is the change of the sign of the constant $I$.
More specifically:

\mbox{} \\
{\bf Patch ``+''~:} In the patch $r>0$ we choose the positive 
sign, and put $c=1$ (which is the standard choice), so that $X>X_0$, 
and therefore we get
\begin{align}
      X=I^{-1}r+X_0, \quad r>0.
\end{align}

\mbox{} \\
{\bf Patch ``--''~: } In the patch $r<0$, using the ambiguity in the 
definition of the integrating factor $I$, we choose the negative 
sign, i.e.~$c=-1$ in~(\ref{eq:I}). In this way also in this 
patch the dilaton is positively defined, $X>X_0$, namely
\begin{align}
      X=-I^{-1}r+X_0, \quad r<0.
\end{align}

\mbox{} \\
Let us moreover remark that non-smooth potentials may lead to
line elements which are smooth as a function of the dilaton (case~(i)), 
but non-smooth with respect to the 'physical' coordinate, because
of the non-smooth relation between $X$ and $r$. This feature is
already known in dilaton gravity models, see e.g. the ``kink''
discussion in~\cite{Grumiller:2003ad}.
A similar situation occurs with Rosen versus Brinkmann coordinates 
with PP-waves, (cf~e.g.\cite{Aichelburg:1996sg}).

\section{Example: a static point-like source}

In the recent paper of Melis and Mignemi on ``Two-dimensional 
static black holes with point-like sources''~\cite{Melis:2004bk} 
a very special dilaton gravity model is considered, supplemented 
with a matter part $\mathcal{L}_m=2\kappa^2 X m\delta(r-r_0)$. Such a matter 
contribution in the action is covered by the above 
analysis of case (ii).
With the specified potentials $V(X)=-l^2 X^h$ and $U(X)=0$ we find
\begin{align}
      \tilde{I}(X)=
      \begin{cases}
      c\,G_0 & X>X_0,\\
      c & X<X_0.
      \end{cases}
\end{align}
For case (ii) this yields
\begin{align}
      \tilde{w}(X)=
      \begin{cases}
      -l^2 c\, G_0 (X^{h+1}-X_0^{h+1})/(h+1) & X> X_0,\\[1ex]
      -l^2 c\, (X^{h+1}-X_0^{h+1})/(h+1) & X<X_0.
\end{cases}
\end{align}
Choosing the dilaton to be positive, which is equivalent to considering
the two patches described above ($c=\pm 1$) for positive and 
negative $r$ respectively, the Killing norms in the 
respective patches are
\begin{align} 
      K_{\pm}(X)=\pm2G_0\Bigl(\mathcal{C}_{\pm}\pm\frac{G_0l^2}{h+1}
      (X^{h+1}-X_0^{h+1})\Bigr),
\end{align}
where the upper and lower sign correspond to patch $+$ and $-$ respectively.
The continuity of the metric at $X_0$ yields the matching condition
$\mathcal{C}=\pm \mathcal{C}_{\pm}$. Hence, the Killing norm reads
\begin{align} 
      K(X)=2G_0\Bigl(\mathcal{C}+\frac{G_0l^2}{h+1}
      \Bigl(X^{h+1}-X_0^{h+1}\Bigl)\Bigr).
\end{align}
Comparing the latter with the result of~\cite{Melis:2004bk}, where
\begin{align}
      K(X)=\frac{1}{h+1}\left(X^{h+1}-\frac{h+3}{2}
      \left(\frac{\kappa^2m}{l}\right)^\frac{h+1}{h}\right),
\end{align}
simply yields 
\begin{align}
      G_0^2=\frac{1}{2 l^2},\qquad
     \mathcal{C}=-\frac{\sqrt{2}}{4}(\kappa^2m)^{\frac{h+1}{h}}l^{-\frac{1}{h}},
\end{align}
where we put $X_0=(\kappa^2m/l)^{1/l}$.
This shows that the model considered in~\cite{Melis:2004bk}
belongs to the general class of dilaton gravity models
with non-smooth potentials. 

Finally we find that the ADM mass $M_{\rm ADM}$ of~\cite{Melis:2004bk} 
(cf~\cite{Mann:1993yv,Kummer:1995qv,Kummer:1997si})
and the conserved quantity $\mathcal{C}$ are related by
\begin{align}
      M_{\rm ADM} = -\frac{1}{\sqrt{2}}\frac{h+3}{h+1}\mathcal{C}.
      \label{mass}
\end{align}
One has to be careful with~(\ref{mass}) for general $h$.
Strictly speaking the definition of the ADM mass requires asymptotic 
flatness, for $h=0$ the spacetime is indeed flat. But for $h=1$ these
solutions have constant curvature. A canonical mass definition 
in the context of two-dimensional dilaton gravity models can be 
found in the appendix~A of~\cite{Grumiller:2004wi}, where the 
relation between the conserved quantity $\mathcal{C}$ and
the mass for different models is discussed. 

\section{Final remarks and outlook}

We have shown how $\delta$-like terms in the first order
dilaton gravity action can be dealt with. The described method
was applied to a static point-like source term, where it was
possible to recover previous results as a special class of
generic actions with non-smooth potentials.

Let us remark that the effective line element of the virtual black 
hole (VBH)~\cite{Grumiller:2001rg,Grumiller:2002dm} contains 
$\delta$-like contributions, which were not covered in the present
work due to the assumption of a continuous line element. The Ricci
scalar of VBHs contains $\delta$ and $\delta'$ terms which however
yield a vanishing Einstein-Hilbert action.

Non-smooth line elements are also well known in the classical
theory of general relativity, e.g.~thin shells~\cite{Israel:1966rt,Berezin:1987bc}
are not smooth either. The same is also true for constant
density perfect fluid solutions, where the energy-momentum tensor
is not continuous at the boundary of the stellar object and
hence the metric is of type $C^1$, which cannot be 
improved.

The first order formalism can also be applied to brane world
scenarios, as was already outlined in the introduction. The 
metric of the simplest brane spacetime~\cite{Brax:2004xh} is given by
\begin{align}
      ds^2 = e^{-2k|y|}\eta_{\mu\nu}dx^{\mu}dx^{\nu}-dy^2,
      \label{eq:brane}
\end{align} 
where the brane is located at $y=0$ and 
$\eta_{\mu\nu}={\rm diag}(+1,-1,-1,\ldots)$. From the dilaton
gravity point of view it is natural to choose the dilaton to be
$X(y)=\exp(-k|y|)$, where the non-smoothness of the metric is
due to the relation between the dilaton and the two-dimensional 
coordinate (motivated by the above case (ii)).
With the potentials $U=-1/X$ and $V=0$ in action~(\ref{eq:note1})
and taking the negative sign of the integrating factor for $y>0$
and the positive one for $y<0$ into account, we recover the 
above metric~(\ref{eq:brane}) from the general 
solution~(\ref{eom1})--(\ref{eom4}). 

More general brane worlds are those, where for example the geometry 
induced on the brane is homogeneous and isotropic, hence compatible 
with the cosmological principle~\cite{Brax:2004xh}, or where the 
induced geometry is spherically symmetric, e.g.~\cite{Harko:2004ui,Mak:2004hv}. 
These can also be analysed with the methods 
described if the dilaton potentials are chosen appropriately.

We hope that the present formulation will be of help in particular
in the analysis of dynamical models like the scattering of a
black hole and a brane or a dynamical many brane scenario, where
for example two branes collapse.

\section*{Acknowledgement}
We deeply thank D.~Grumiller for suggestions regarding the manuscript 
and for useful discussions. Moreover we thank H.~Balasin and 
D.~Vassilevich for the useful discussions.

The work of CGB was supported by the Junior Research Fellowship of The
Erwin Schr\"odinger International Institute for Mathematical Physics.

\end{document}